\documentstyle[12pt]{article}
\newcommand{\NN}{\nonumber}
\renewcommand{\theequation}{\arabic{section}.\arabic{equation}}
\newcommand\p{^\prime}
\newcommand\pp{^{\prime\prime}}
\def\ts{\thinspace}

\newcommand\wh{\widehat}
\newcommand\ccon{^*}
\newcommand\beq{\begin{equation}}
\newcommand\eeq{\end{equation}}
\newcommand{\rref}[1]{(\ref{#1})}
\newcommand\bear{\begin{array}}
\newcommand\enar{\end{array}}
\newcommand\Bear{\begin{eqnarray}}
\newcommand\Enar{\end{eqnarray}}
\newcommand\Bears{\begin{eqnarray*}}
\newcommand\Enars{\end{eqnarray*}}
\newcommand\lang{\langle}
\newcommand\llang{\langle\!\langle}
\newcommand\rang{\rangle}
\newcommand\rrang{\rangle\!\rangle}

\def\ASPM{{\it Advanced Studies in Pure Mathematics\ts}}
\def\CMP{{\it Comm.\ts Math.\ts Phys.\ts}}

\def\IJMP{{\it Int.\ts J.\ts Mod.\ts Phys.\ts}}

\def\NP{{\it Nucl.\ts Phys.\ts}}
\def\PL{{\it Phys.\ts Lett.\ts}}

\def\PTP{{\it Prog.\ts Theor.\ts Phys.\ts}}

\def\Zm{Zamolodchikov}
\def\AZm{A.\ts B.\ts \Zm}
\def\dur{H.\ts W.\ts Braden, E.\ts Corrigan, P.\ts E.\ts Dorey and R.\ts
Sasaki}
\begin{document}
\begin{flushleft}
{\it YITP Uji Research Center}
\begin{flushright}
YITP/U-95-22 \\
June 1995\\
hep-th/9507001
\end{flushright}
\end{flushleft}

\begin{center}
\vspace*{1.0cm}

{\LARGE{\bf  Instability of Solitons in imaginary coupling affine Toda Field
 Theory }}

\vskip 1.5cm

{\large {\bf S. Pratik  KHASTGIR
\footnote{Monbusho Fellow.}
and Ryu SASAKI
\footnote{Supported partially by the grant-in-aid for Scientific
Research,
Priority Area 231 ``Infinite Analysis'' and General Research (C) in Physics,
Japan Ministry of Education.}}}

\vskip 0.5cm

{\sl Uji Research Center, Yukawa Institute for Theoretical Physics,} \\
{\sl Kyoto University,} \\
{\sl Uji 611, Japan.}

\end{center}

\vspace{1 cm}

\begin{abstract}
Affine Toda field theory with a pure imaginary coupling constant
is a non-hermitian theory.
Therefore the solutions of the equation of motion are complex.
However, in $1+1$ dimensions
it has many soliton solutions with remarkable properties, such as real
total energy/momentum and mass.
Several authors calculated quantum mass corrections of the solitons
by claiming these solitons are stable. We show that there exists a
large class of classical solutions which develops singularity
after a finite lapse of time. Stability claims, in earlier literature,
were made ignoring these solutions.
Therefore we believe that a formulation of quantum theory on a firmer
basis is necessary in general and for the quantum mass corrections of
solitons, in particular.

\end{abstract}

\vspace{1 cm}

\baselineskip=12pt 
\section{Introduction}
\setcounter{equation}{0}

In this paper we will address the problem of the stability of solitons
in the imaginary coupling affine Toda field theory, which is obtained from
the affine Toda field theory by replacing the real coupling constant
$\beta$ by a purely imaginary one $i\beta$, $i=\sqrt{-1}$.
The affine Toda field theory is one of the best understood field theories
at the classical \cite{MOPa}\ts and at the quantum levels
\cite{AFZa,BCDSa,BCDSc,CMa,DDa}\cite{DGZc,CDS}, thanks to its integrability.
It is the close connection between the affine Toda field theory and
the conformal field theory in 2 dimensions,
another group of best understood quantum field theories,
 (integrable deformation of conformal field theory \cite{ZEYSY,HMBL})
that led to the interesting but controversial ``imaginary coupling" affine
Toda field theory.

This apparently tiny change $\beta\to i\beta$ brings  huge differences between
the
affine Toda field theory and its imaginary coupling counterpart.
Among them the following two aspects are most prominent:
The first is the emergence of soliton solutions and other interesting
exact solutions in
the imaginary coupling theory,
just as in the well known sine-Gordon theory,
which is the simplest example of the imaginary coupling affine Toda field
theories.
In contrast, the affine Toda field theory is known to have no solitons.
The second is the lack of reality/hermiticity of the Lagrangian and action
in all the imaginary coupling theories except for the sine-Gordon theory.

Many interesting and beautiful results on solitons have been
obtained by various authors.
By applying Hirota's method, Hollowood \cite{Hoa}\ts
obtained various simple soliton
solutions in the
$a_n^{(1)}$ theory.
The total energies of these one soliton solutions are real, although
the solutions themselves and the energy densities are complex.
The masses of the solitons are found to be proportional to the masses
of the fundamental particles of the corresponding affine Toda field theory.
This result was further developed by many authors
\cite{Hob,MM,OTUa,Naa,KO,Und}.
A complete set of soliton solutions was obtained by invoking
representation theory of affine Lie algebras by Olive and collaborators
\cite{OTUa,KO}.
The mass spectrum of the one soliton solutions is now known and
it is related to the mass spectrum of the fundamental particles in
the real coupling theories in a very interesting way.

\medskip

Hollowood \cite{Hoc}\ts then set a new trend by calculating
``quantum mass corrections" to the solitons
in $a_n^{(1)}$ theory.
It was reported that the soliton mass ratios in the $a_n^{(1)}$
theories were unchanged after one-loop corrections.
Then ``quantum mass corrections" to the solitons in all the affine Toda
theories were also investigated by Watts \cite{Waa}, Delius and Grisaru
\cite{DGb}\ts and MacKay and Watts \cite{MW}.
They obtained similar  but slightly differing results
and the relationship  between the classical soliton masses and the  quantum
mass corrections seemed to be more involved.

On the other hand,
the lack of reality/hermiticity of the Lagrangian does not seem
to have attracted much attention.
This is a rather strange situation, since the hermiticity or reality
is sacrosanct in any physical theory and especially in quantum physics,
in which non-hermitian Lagrangian or Hamiltonian implies non-unitarity
and non-conservation of probability.
The classical theory is well defined mathematically, even if the Lagrangian is
non-hermitian.
Although the physical interpretation of the solutions of the equation of motion
is dubious, the concept of solutions is solid.
In contrast, to the best of our knowledge, the quantum field theory or
even quantum mechanics of non-hermitian systems simply does not exist.
Thus, at present, quantum  affine Toda field theory with imaginary coupling
should be considered to be of heuristic nature.
Therefore the clarification of the hermiticity issue is essential
 for any serious treatments of the quantum solitons,
especially for those of quantum mass corrections.
The lack of unitarity in quantum field theory is usually related with the
lack of stability of the solutions of the corresponding classical
field theory, which are the solitons in the present case.

In this context, certain stability arguments of
soliton solutions connected with the ``twisted reality" or
``twisted hermiticity" relations were produced by
Hollowood \cite{Hoc}, Evans \cite{Eva}\ts and Delius and Grisaru \cite{DGb}.
We will show that none of these arguments is satisfactory.
We also give various explicit solutions
which develop singularity after a finite lapse of time.
Moreover, these singular solutions  are far more abundant than the
non-singular ones.
To be more precise,  the moduli  space of the singular 1-soliton solutions
is two dimensional whereas  the non-singular 1-soliton solutions
have one dimensional moduli  space.
These results cast a big question mark on
the works of quantum mass corrections of solitons, in particular and on those
of the
soliton physics/mathematics in
the imaginary coupling affine Toda field theory in general.
The problem of the hermiticity and/or unitarity
in  the imaginary coupling affine Toda field theory
deserves a far greater attention than has been given,
since the stake is very high.

This paper is organised as follows:
in section 2 we briefly review the essentials of affine Toda field
theory in order to set the stage and to introduce  notation.
In section 3 derivation of the simple solitons are given and
some of their salient properties are recapitulated.
In Section 4 various existing stability arguments of solitons are examined
and shown to have flaws.
In Section 5 and 6 instability of ``vacuum" and single soliton solutions
are shown by examining the time evolution of
the explicit solutions.
The moduli space of generalised single-soliton solutions
are also introduced in order to show that the generic solutions are unstable.
Section 7 is devoted to a brief summary.
Appendix A gives a simple example of $2\times2$ matrix satisfying
the ``twisted reality condition" but failing to have real eigenvalues.
Appendix B gives another simple example of $2\times2$ symmetric matrices
which are not hermitian. It is shown that its eigenvectors are
not guaranteed to span the entire vector space.


\section{Affine Toda field theory}
\setcounter{equation}{0}

Affine Toda field
theory \cite{MOPa} is a
massive scalar field theory with exponential interactions in $1+1$
dimensions
described by the Lagrangian
\begin{equation}
{\cal L}={1\over 2}
\partial_\mu\phi^a\partial^\mu\phi^a-V(\phi ),
\label{ltoda}
\end{equation}
in which the potential is given by
\begin{equation}
V(\phi )={m^2\over
\beta^2}\sum_0^rn_je^{\beta\alpha_j\cdot\phi}.
\label{vtoda}
\end{equation}
The field $\phi$ is an $r$-component scalar field, $r$ is the rank of a
compact semi-simple Lie algebra $g$ with $\alpha_j$;
$j=1,\ldots,r$ being its simple roots. The roots are normalised so that long
roots have length 2, $\alpha_L^2=2$. An additional root,
$\alpha_0=-\sum_1^rn_j\alpha_j$ is an integer linear combination of the simple
roots, is called the affine root;
 it corresponds to the extra spot on an extended Dynkin diagram
for $\hat g$ and $n_0=1$.
When the term containing the extra root is removed, the theory becomes
conformally invariant (conformal Toda field theory).
The simplest affine Toda field theory, based on the simplest Lie algebra
$a_1^{(1)}$, the algebra of ${\wh su(2)}$,
is called sinh-Gordon theory, a cousin of
the well known sine-Gordon theory. $m$ is a real parameter setting
the mass scale of the theory and $\beta$ is a real coupling constant,
which is relevant only in quantum theory.

Toda field theory is integrable at the classical level due to
the presence of an infinite number of conserved quantities.
Many beautiful properties of Toda field theory, both at the classical
and quantum levels, have been uncovered in recent years.
In particular, it is firmly believed that the integrability survives
quantisation.
The exact quantum S-matrices are known
\cite{AFZa,BCDSa,BCDSc,CMa,DDa},\cite{DGZc,CDS}
 for all the
Toda field theories based on non-simply laced algebras as well as those
based on simply laced algebras.
The singularity structure of the latter S-matrices, which in some
cases contain poles up to 12-th order \cite{BCDSc},
is  beautifully explained in terms of the
singularities of the corresponding Feynman diagrams
\cite{BCDSe}, so called Landau singularities.

The imaginary coupling affine Toda field theory is obtained simply
by replacing $\beta$ by $i\beta$ ($i=\sqrt{-1}$) in the Lagrangian
\begin{equation}
{\cal L}_I={1\over 2}
\partial_\mu\phi^a\partial^\mu\phi^a- V_I(\phi ),\quad V_I(\phi )=-{m^2\over
\beta^2}\sum_0^rn_j\left(e^{i\beta\alpha_j\cdot\phi}-1\right),
\label{imtodalag}
\end{equation}
and by reinterpreting the fields $\phi$ as {\em complex}.
The Lagrangian is {\it not hermitian} except for the $a_1^{(1)}$ theory
($r=1$, $n_1=n_0=1$, $\alpha_1=-\alpha_0=\sqrt2$) with a real field.
In the rest of this paper,
 $a_1^{(1)}$ theory is excluded.
The  equation of motion obtained from the above Lagrangian reads
\beq
\partial_\mu^2\phi=-{m^2\over{i\beta}}\sum_0^rn_j\alpha_j
e^{i\beta\alpha_j\cdot\phi}.
\eeq
It is easy to see that it has no real solutions
except for the trivial constant solutions corresponding to the ``minima"
of the imaginary potential\footnote{Points in \rref{potminima}\ts
simply correspond to the stationary points of the potential $V_I$,
${\partial V_I\over{\partial\phi}}=0$.
The second derivative matrix
${\partial^2V_I\over{\partial\phi^a\partial\phi^b}}=m^2\sum
n_j\alpha_j^a\alpha_j^b$, which is equal to the classical $({\rm mass})^2$
matrix, is positive definite.
Since the potential is an analytic function
of the field $\phi$, it has neither a minimum nor a maximum in any open domain.
In other words the potential $V_I$ as a function of the complex field
$\phi$ is {\it not bounded from below}, another sign of instability.}:
\beq
{2\pi\over\beta}\sum_{j=1}^rk_j\lambda_j,\quad k_j:~{\rm integer}.
\label{potminima}
\eeq
Here $\{\lambda_j\}$, $j=1,\ldots,r$ is the dual basis to the simple roots
$\{\alpha_j\}$:
\beq
\alpha_j\cdot\lambda_k=\delta_{jk}.
\label{dualbas}
\eeq

\medskip

It should be noticed that the structure of the Lax pair, the existence of
an infinite set of conserved quantities in involution, the corner stone of the
integrability,
and etc. etc are the same in the imaginary coupling theory as in the real
coupling theory \cite{Hoc,Eva}.
However, their actual contents are markedly different.
In the real coupling theory, the conserved energy is {\it positive definite},
namely each term is positive.
Thus if one follows the time evolution of a regular initial data $\phi(x,0)$,
$\partial_t\phi(x,0)$ with finite energy,
the field $\phi(x,t)$ and its first derivatives are
always finite everywhere,
since any singularity would violate the conservation of energy.
In the imaginary coupling theory, the conservation of energy still holds
but the energy has negative as well as positive terms.
The conservation of energy fails to prevent the singularities in
the time evolution.
We show this phenomenon by explicit examples in section 5 and 6.
Other simple examples of singularities in the case of
non-positive definite energy
caused by integrable boundary interactions were given in Ref.\cite{FSa}.

\medskip

In the rest of this paper we will discuss the $a_n^{(1)}$ theory only
($r=n$, $n_j=1$, $j=0,1,\ldots,n$) for definiteness and simplicity.
This also makes it easy to concentrate on
the fundamental and universal problems of (non) hermiticity and
stability without being bothered by the Lie algebra technicalities.

\section{Solitons}
\setcounter{equation}{0}

In this section we recapitulate some of the results on the explicit
soliton solutions which are necessary for our purposes.
Only the very fundamental features of the soliton solutions are relevant here,
so we follow the elementary method of Hollowood \cite{Hoa}.
Like the sine-Gordon solitons, these solitons interpolate various
``vacua" or the ``minima" of the complex potential
\rref{potminima}.
We start from the following Hirota ansatz
\beq
\phi(x,t)=-{1\over{i\beta}}\sum_{j=0}^n
\alpha_j\log\tau_j.
\label{Hiranz}
\eeq
In terms of $\tau_j$ the equation of motion can be decoupled
into
\beq
{\ddot \tau}_j\tau_j-{\dot \tau}_j^2-\tau\pp_j\tau_j+{\tau\p_j}^2
=m^2(\tau_{j-1}\tau_{j+1}-\tau_j^2).
\label{taueq}
\eeq
The label on $\tau_j$ is understood modulo $n+1$, which reflects
the periodicity of the $a_n^{(1)}$ Dynkin diagram.

One soliton solution is obtained by assuming
\beq
\tau_j=1+\tau_j^{(1)}.
\label{onesolanz}
\eeq
By substituting \rref{onesolanz}\ts into \rref{taueq}\ts we get
\Bear
{\ddot \tau}_j^{(1)} &-&{\tau\pp_j}^{(1)} -m^2(\tau_{j-1}^{(1)}+
\tau_{j+1}^{(1)}-2\tau_j^{(1)})\NN \\
&+&{\ddot \tau}_j^{(1)}\tau_j^{(1)}-({\dot \tau}_j^{(1)})^2
-{\tau\pp_j}^{(1)}\tau_j^{(1)}+({\tau\p}_j^{(1)})^2
-m^2(\tau_{j-1}^{(1)}\tau_{j+1}^{(1)}-(\tau_j^{(1)})^2)=0.
\label{onesoleq}
\Enar
A characteristic feature of the Hirota method is that this
equation is decomposed into linear and quadratic parts:
\Bear
{\ddot \tau}_j^{(1)} -{\tau\pp_j}^{(1)} -m^2(\tau_{j-1}^{(1)}+
\tau_{j+1}^{(1)}-2\tau_j^{(1)})&=&0,
\label{linoneeq} \\
{\ddot \tau}_j^{(1)}\tau_j^{(1)}-({\dot \tau}_j^{(1)})^2
-{\tau\pp_j}^{(1)}\tau_j^{(1)}+({\tau\p}_j^{(1)})^2
-m^2(\tau_{j-1}^{(1)}\tau_{j+1}^{(1)}-(\tau_j^{(1)})^2)&=&0,
\label{quadoneeq}
\Enar
and the quadratic part is always satisfied
by the solution of the linear part.
The linear equation is solved by
\beq
\tau_j^{(1)}=\exp(\sigma x-\lambda t+x_0+j\rho),
\label{onesolgensol}
\eeq
for constants $\sigma$, $\lambda$, $x_0$ and $\rho$.
The periodicity in the label $j$ on $\tau_j$ then implies
\beq
\rho={2\pi ia\over{n+1}},\qquad a:~{\rm integer}\quad 1\leq a\leq n.
\label{rhoform}
\eeq
The parameters $\sigma$, $\lambda$ and the integer $a$
are constrained by
\beq
{\cal F}(\sigma,\lambda,a)\equiv\sigma^2-
\lambda^2-4m^2\sin^2{\pi a\over{n+1}}=0,
\label{onsellcond}
\eeq
in order to satisfy \rref{linoneeq}.
It is very easy to verify that the quadratic equation is also satisfied.

\medskip

Thus we arrive at the explicit one soliton
solution
\beq
\phi_a(x,t)=-{1\over{i\beta}}\sum_{j=0}^n\alpha_j\log
[1+\omega^{aj}\exp(\sigma x-\lambda t+x_0)],
\label{1solgen}
\eeq
in which the parameters $\sigma$ and $\lambda$ should satisfy
\beq
\sigma^2-\lambda^2=4m^2\sin^2{\pi a\over{n+1}},\label{onshell}
\eeq
and $\omega$ is a primitive root of unity $\omega=e^{2\pi i\over{n+1}}$,
$\omega^{n+1}=1$.
The right hand side of \rref{onshell}\ts is is simply the ${\rm mass}^2$ of the
fundamental particles in $a_n^{(1)}$ Toda field theory.
The above 1-soliton solutions are classified into three different types
as follows:
\beq
\begin{array}{ll}
1)~ \sigma, \lambda : {\rm Real} &{\rm r-soliton} \\
2)~ \sigma, \lambda : {\rm pure~ Imaginary} &{\rm i-soliton} \\
3)~ \sigma, \lambda : {\rm Complex} &{\rm c-soliton}.
\end{array}
\label{3typesol}
\eeq
The c-solitons contain all the 1-soliton solutions not belonging to the
r-solitons and i-solitons; for example,
$\sigma$: real  and $\lambda$: pure imaginary.
The parameters $(\sigma,\lambda)$ have one real degree of freedom in the cases
of
r-solitons and i-solitons, whereas $(\sigma,\lambda)$ have two real
degrees of freedom in the c-soliton case.
The r- and i-solitons are located at the boundaries of the moduli space of the
c-soliton solutions.
As we will see in section 5, these three types of 1-soliton solutions
have very different characters.
The other parameter $x_0$ is in general complex and
its real part  $x_{0R}$ is related to the location of the
soliton at $t=0$. The imaginary part of $x_0$, $x_{0I}$ can be restricted to
$0\leq x_{0I}<2\pi$ without loss of generality.

\medskip

In the rest of this section we give the two soliton
solutions without derivation \cite{Hoa}.
Let us define
\beq
y^{(p)}_j=\sigma_px-\lambda_pt+x^{(p)}_0+{2\pi ia_p\over{n+1}}j,
\label{ypdef}
\eeq
in which the parameters $\sigma_p$, $\lambda_p$ and the integer $a_p$ satisfy
the constraint
$$
{\cal F}(\sigma_p,\lambda_p,a_p)=0.
$$
Then a general two soliton solution is given by
\beq
\tau_j=1+e^{y^{(1)}_j}+e^{y^{(2)}_j}+e^{y^{(1)}_j+y^{(2)}_j+\gamma_{(12)}},
\label{twosoltau}
\eeq
in which the interaction function $e^{\gamma_{(pq)}}$ is given by \cite{Hoa}
\beq
e^{\gamma_{(pq)}}=-{{\cal F}(\sigma_p-\sigma_q,\lambda_p-\lambda_q,a_p-a_q)
\over
{{\cal F}(\sigma_p+\sigma_q,\lambda_p+\lambda_q,a_p+a_q)}}.
\label{intfunc}
\eeq

\section{``Stability" of 1 Soliton Solutions}
\setcounter{equation}{0}

As a ``prerequisite" for calculating ``quantum mass corrections"
of solitons, Hollowood \cite{Hoc}, Evans \cite{Eva}\ts and Delius and Grisaru
\cite{DGb}\ts produced
certain arguments that the soliton solutions are classically
``stable".
In this section we recapitulate the essence of their ``stability" arguments
and show that these are flawed.
They picked up a stationary r-soliton solution located at the origin
\beq
{\bar \phi}_a(x)=-{1\over{i\beta}}\sum_{j=0}^n\alpha_j
\log[1+\omega^{aj}e^{m_ax}],\quad m_a=2m\sin{\pi a\over{n+1}},
\label{stasol}
\eeq
and considered a small perturbation around it:
\beq
\phi(x,t)={\bar \phi}(x)+\eta(x,t).
\label{smallpert}
\eeq
 From the equation
of motion for $\phi$, a linearised equation for
the small perturbation $\eta$ was derived
\beq
\partial_\mu^2\eta +m^2\sum_{j=0}^n\alpha_j(\alpha_j\cdot\eta)
\exp(i\beta\alpha_j\cdot{\bar\phi})=0.
\label{etalineq}
\eeq
Assuming simple time dependence
$$
\eta(x,t)=\eta(x)e^{i\nu t},
$$
the linearised equation of motion was reduced to an eigenvalue
problem
\beq
{\cal D}\eta=\nu^2\eta,
\label{eigprob}
\eeq
in which ${\cal D}$ was a {\it non-hermitian}
second order differential operator of the following form
\beq
{\cal D}=-{d^2\over{dx^2}}+m^2\sum_{j=0}^n\alpha_j\otimes\alpha_j
\exp(i\beta\alpha_j\cdot{\bar\phi}).
\label{ddef}
\eeq
They argued that ``if the spectrum of ${\cal D}$ --for bounded
 eigenfunctions--
was real and positive; hence, the frequencies $\nu$ were real,
then the small perturbations to ${\bar\phi}$ would not diverge".

\medskip
If ${\cal D}$ is hermitian, then obviously its eigenvalues
$\nu^2$ are real, the eigenfunctions belonging to different eigenvalues
are orthogonal to each other and they constitute a complete basis
of the function space.
However, ${\cal D}$ {\it is non-hermitian}.
Its eigenvalues are in general complex and
the eigenfunctions are not guaranteed to form a complete orthogonal basis
of the entire function space.
Therefore, the ``stability argument" based on the eigenfunctions of ${\cal D}$
is in general incomplete.

\medskip
Hollowood and Evans' \cite{Hoc,Eva}\ts argument that ``$\nu^2$ are real" goes
as follows:
First ${\bar\phi}$ satisfies the following ``twisted reality" condition
\beq
{\bar\phi}\ccon(x)=-M{\bar\phi(x)},\quad *~{\rm denotes~complex~conjugation},
\label{solrealitycond}
\eeq
in which $M$ acts as a $Z_2$ symmetry of the roots
\beq
M\alpha_j=\alpha_{n+1-j},\quad \alpha_{n+1}\equiv\alpha_0,
\label{Mdefs}
\eeq
and it also satisfies the conditions
\beq
\qquad M^2=1,\quad M^t=M.
\label{Mconds}
\eeq
 From this it follows that
\beq
{\cal D}^\dagger=M{\cal D}M,\qquad \dagger~{\rm denotes~hermitian~conjugation}.
\label{hermproD}
\eeq
Hence, they argue that ``${\cal D}$ is hermitian" with respect to the following
``inner product":
\beq
\lang f,g\rang=\int_{-\infty}^\infty dx\ts f^\dagger(x)\cdot Mg(x).
\label{newinprodef}
\eeq
In fact, it is easy to see
\beq
\lang f,{\cal D}g\rang=\int_{-\infty}^\infty dxf^\dagger(x)\cdot M{\cal D}g(x)=
\int_{-\infty}^\infty dxf^\dagger(x)\cdot{\cal D}^\dagger Mg(x)=\lang{\cal
D}f,g\rang.
\label{hermproDveri}
\eeq
Based on this they assert that the spectrum of ${\cal D}$ is real.

\medskip
However, {\it the flaw lies in the point that} $\lang f,g\rang$
{\it does not define an inner product} since it is not {\it positive definite}.
Namely, $\lang f,f\rang$ can be positive, zero or negative and $\lang
f,f\rang=0$ does not imply
$f=0$.
Supposing that $f$ is an eigenfunction of ${\cal D}$ with eigenvalue $\nu^2$,
then we can calculate $\lang f,{\cal D}f\rang$ in two ways to obtain
\beq
\left(\nu^2-(\nu^2)\ccon\right)\lang f,f\rang=0.
\label{nonrealrel}
\eeq
But we cannot conclude from this that $\nu^2$ is real:
$$
\nu^2\neq(\nu^2)\ccon,\quad{\rm in~general},
$$
because of the possibility of $\lang f,f\rang=0$, see \rref{zeronorm}.
The fact that the ``inner product" $\lang f,g\rang$ is not positive definite
can be easily seen when one takes a basis of the n-dimensional vector space
(the Cartan subalgebra of $a_n$) such that $M$ is diagonal.
Since $M^2=1$, $M$ has eigenvalue $\pm1$ subspaces and the $-1$ subspace
violates the positive definiteness.
In appendix A we give a simple example of $M$ and
${\cal D}$ satisfying the conditions
\rref{Mconds}\ts and \rref{hermproD},
but ${\cal D}$ failing to produce real eigenvalues,
or failing to give a complete orthogonal basis consisting of its eigenvectors.

\medskip
Delius and Grisaru's argument \cite{DGb}\ts is slightly different.
They assumed a {\it complete set of orthonormal eigenfunctions}
$\eta_k(x)$ of ${\cal D}$
\beq
{\cal D}\eta_k(x)=\nu_k^2\eta_k(x),\quad
\llang\eta_k,\eta_{k\p}\rrang=\delta_{kk\p},
\label{DGeig}
\eeq
with respect to an ``inner product" without complex conjugation
\beq
\llang f,g\rrang=\int_{-\infty}^\infty dx f^t(x)\cdot g(x),
\qquad t~{\rm denotes~transpose}.
\label{DGinerpro}
\eeq
They argued that ${\cal D}$ was not hermitian but symmetric
$$
\llang f,{\cal D}g\rrang=\llang{\cal D}f,g\rrang.
$$
However, $\llang f,g\rrang$ cannot define an inner product, since $\llang
f,f\rrang$ is neither
real nor positive.
Therefore, the ``stability analysis" and ``quantisation" based on a
``complete set" of eigenfunctions of ${\cal D}$ cannot be justified.
In appendix B we give a simple example of $2\times2$ matrix
which is symmetric but non-hermitian and show that
its eigenvectors need not form a complete orthogonal basis.

\bigskip
Further they went on calculating the explicit eigenfunctions of
${\cal D}$ with real eigenvalues for the evaluation of the
``quantum corrections" to the masses
of solitons.
Such ``quantisation procedure" is not well founded because
the eigenfunctions do not span the complete
function space. 
For, if we assume that ${\cal D}$ has real eigenvalues only and that
the corresponding eigenfunctions form a complete orthogonal basis,
then we can easily prove
that ${\cal D}$ is hermitian, which is a contradiction.


\section{Blowing up Solutions 1}
\setcounter{equation}{0}

In the previous section we have confirmed that the linear operator ${\cal D}$
\rref{ddef}\ts describing small perturbations around a stationary r-soliton
solution is not hermitian.
Thus the ``small perturbations" always contain certain components which
grow exponentially in time and the linear approximation eventually
breaks down.
In this and subsequent sections we will show that most solutions of the
imaginary coupling
affine Toda field theory  really develop singularities after certain time
and that the theory is {\it unstable}.

Let us first look at the asymptotic ($x\to\pm\infty$)
properties of the three types of 1-soliton solutions \rref{3typesol}\ts at
$t=0$.
All of them are {\it bounded functions of $x$}.
For r- and c-solitons, let us assume that Re\ts$\sigma>0$.
Then at $x\to+\infty$ the expression \rref{1solgen}\ts
can be simplified as
\Bear
-i\beta\phi(x,t)&\approx&\sum_{j=0}^n\alpha_j\log[\exp(\sigma x+x_0
+{2\pi ia\over{n+1}}j)]\NN \\
&=&\sum_{j=0}^n\alpha_j\left(\sigma x+x_0
+{2\pi ia\over{n+1}}j\right)\NN \\
&=&\sum_{j=0}^n\alpha_j
\left({2\pi ia\over{n+1}}j\right).
\label{1solasymp}\\
\Enar
The parts proportional to $\sigma x+x_0$ cancel with each other due to the
relation of the
simple roots
$\sum_{j=0}^n\alpha_j=0$.
At $x\to-\infty$ the r- and c-soliton solutions simply go to zero.
Similar arguments can be made for the i-soliton solutions.
It is easy to show that any combination of the above soliton solutions
shares this property.

In this section we will show that all c-soliton
solutions develop singularities after a finite time.
The singularity is caused by vanishing of the arguments of the logarithms.
Let us follow the time developments from the `initial' time $t=0$.
At $t=0$ the argument of the $j$ th logarithm is
\beq
1+\exp(\sigma x+x_0+{2\pi ia\over{n+1}}j).
\label{jargum}
\eeq
In order this to vanish $x$ must be a root of
\beq
\sigma x+x_0+{2\pi ia\over{n+1}}j=i(2m+1)\pi,\quad m:{\rm integer}
\label{vancon}
\eeq
However, this equation does not have a real root in general.
If it has one, we could have started with a slightly different $x_0$ and
$\sigma$.
To sum up,
$$
{1\over\sigma}\left(i(2m+1)\pi -{2\pi ia\over{n+1}}j-x_0\right)
$$
is in general complex, unless $x_0$ and $\sigma$ are fine tuned.
So $\phi(x,0)$ is {\it regular everywhere}. From the continuity of
\rref{1solgen} $\phi(x,t)$ is {\it regular everywhere} for small enough $t$.

\medskip

As $t$ increases from zero, the argument of the c-soliton
changes. The condition for vanishing argument \rref{vancon}\ts now reads
\beq
\sigma x-\lambda t+x_0+{2\pi ia\over{n+1}}j=i(2m+1)\pi,\quad m:
{\rm  integer}.
\label{vancont}
\eeq
By introducing the real and imaginary parts of $\sigma$, $\lambda$ and $x_0$
$$
\sigma=\sigma_R+i\sigma_I,\quad \lambda=\lambda_R+i\lambda_I,\quad
x_0=x_{0R}+ix_{0I},
$$
\rref{vancont}\ts can be rewritten as
\beq
\pmatrix{\sigma_R &-\lambda_R\cr
\sigma_I &-\lambda_I\cr}\pmatrix{x \cr t\cr}=\pmatrix{-x_{0R} \cr
(2m+1)\pi-{2\pi a\over{n+1}}j-x_{0I}\cr},
\label{mateqone}
\eeq
which has always a real root, unless
\beq
{\rm det}\pmatrix{\sigma_R &-\lambda_R\cr
\sigma_I &-\lambda_I\cr}=-\sigma_R\lambda_I+\sigma_I\lambda_R=0.
\label{vandet}
\eeq
The above condition means that
\beq
{\sigma_I\over{\sigma_R}}={\lambda_I\over{\lambda_R}},\qquad {\rm or}\quad
\sigma=k\lambda,\quad k:{\rm real}.
\label{degecase}
\eeq
In this case the  condition
$\sigma^2-\lambda^2=4m^2\sin^2{\pi a\over{n+1}}$ \rref{onshell}\ts
can never be satisfied by complex $\sigma$ and $\lambda$.
So we need not worry about the above situation in the case of c-solitons.
The above result also shows  that the r-soliton and the i-soliton solutions
are essentially singularity free, since for them the determinant
\rref{vandet}\ts vanishes.

Let us choose among the  roots of \rref{mateqone}\ts for
various $j$ and $m$, the one having the smallest $|t|$ and call it
$t_M$.
If $t_M<0$ then we change $\lambda\to-\lambda$ and get $t_M>0$.
Therefore we have shown that the c-soliton solutions
always develop  singularity as time increases.

\bigskip
Next let us remark that the above $\phi(x,0)$ can be made as small as we wish
within a given finite interval $[-L,L]$.
Suppose $\sigma_R>0$, then by choosing $x_{0R}$
sufficiently large and negative,
we can make
\beq
|\exp(\sigma x+x_0+{2\pi ia\over{n+1}}j)|<\epsilon \quad {\rm for}\quad |x|<L.
\label{smallamp}
\eeq
In fact
$$
{\rm Max}|\exp(\sigma x+x_0+{2\pi ia\over{n+1}}j)|=e^{\sigma_RL+ x_{0R}},
$$
so that we have to choose $x_{0R}$ such that
$$
e^{\sigma_RL+x_{0R}}<\epsilon=e^{\log\epsilon}.
$$
That is
\beq
x_{0R}<-\sigma_RL+\log\epsilon.
\label{sigmaval}
\eeq

Due to the fact that the influence of the c-soliton solution
on $\phi(x,0)$
can be made as small as we wish in any finite region,
the above result (blowing up of a c-soliton solution)  can also be regarded as
the ``instability" of the
``vacuum", $\phi(x,t)\equiv0$.

This instability can be naively ``understood" if we approximate
\rref{1solgen}\ts
in the region $x<L$,
\beq
-i\beta\phi(x,t)=\sum_{j=0}^n\alpha_j\ts e^{\sigma x-\lambda t +x_0+
{2\pi ia\over{n+1}}j}
\label{1solexp}
\eeq
which has $e^{-\lambda t}$, an exponentially growing or decaying
factor for complex $\lambda$.
It should be noted, however, that this approximation is not valid for $x>L$.

\medskip

Before concluding this section let us remark on the vacuum on which quantum
states should be built.
In any Lorentz invariant quantum field theory, the vacuum is a classical
configuration
(namely a solution of the equation of motion) satisfying
the following two conditions:
1) time and space translational invariance.
2) having the lowest energy (which can be chosen to be zero).
In the affine Toda field theory with real coupling it is   $\phi(x,t)\equiv0$
and unique.

But the situation is very different in imaginary coupling theory.
There is no classical solution satisfying these two conditions.
In short there is no stable vacuum.
Firstly the candidates satisfying the condition 1)
are infinite in number, that is the points \rref{potminima}.
Therefore we call them ``vacua".
The degeneracy of ``vacua" usually indicates
instability in ordinary quantum field theory context.

Secondly, there are infinitely many solutions of
equation of motion which have lower energies than the
$\phi(x,t)\equiv0$ (or ${2\pi\over\beta}\sum k_j\lambda_j$), configuration.
Therefore these ``vacua"  are unstable since they
will decay into lower energy states by quantum tunneling.
Suppose $\phi_0$ is a constant such that $V_I(\phi_0)=-v<0$.
Then consider the solution of the initial value problem
\beq
\phi(x,0)=\phi_0,\quad \partial_t\phi(x,0)=0,
\label{candvac1}
\eeq
which has energy lower than 0.
In this construction the total energy is in fact minus infinity,
$E=-v\times{\rm space volume}$.

There are also solutions having finite negative energy.
Consider the solution of the following initial value problem:
\beq
\phi(x,0)=0,\quad {\rm everywhere}\qquad \partial_t\phi(x,0)=if(x),
\label{candvac2}
\eeq
in which $f(x)$ is a real function and finite everywhere and square integrable
$$
\int_{-\infty}^\infty f(x)^t\cdot f(x)dx=F>0.$$
which has a negative total energy
$$
E=-{1\over2}F<0.$$
None of these negative energy solutions are time and space translational
invariant. We do not know if the solutions of the initial value problems
\rref{candvac1},\rref{candvac2}\ts reman finite or not.
The existence of these negative energy states is another
evidence of the instability of ``vacua" and it gives another difficult hurdle
for constructing the quantum  field theory, if any.


\section{Blowing up Solutions 2}
\setcounter{equation}{0}

Next let us show that any r-soliton solution is unstable
in the same manner as the ``vacuum" is unstable by an addition of
a small c-soliton solution. In other words we show that 2-soliton solutions
consisting of a 1 r-soliton and a 1 c-soliton solutions develop singularity
after a finite time.

For simplicity, let us assume that the r-soliton is at rest near
the origin,
\Bear
\Psi_j^{(a)}&=&1+e^{m_ax+x_0^{(a)}+{2\pi ia\over{n+1}}j}\equiv1+e^{y^a_j},
\label{re1sol}\\
x^{(a)}_{0R}&=&0.
\label{solori}
\Enar
Let us add to it a c-soliton from the right (meaning $\sigma_R>0$),
\Bear
\Psi_j^{(C,b)}&=&1+e^{\sigma x-\lambda t+x_0^{(b)}+{2\pi ib\over{n+1}}j}
\equiv1+e^{y^b_j},
\label{compsol}\\
\sigma^2-\lambda^2&=&4m^2\sin^2{\pi b\over{n+1}},\quad 1\leq b\leq n.
\label{componshell}
\Enar
The total solution is
\beq
-i\beta\phi(x,t)=\sum_{j=0}^n\alpha_j\log[1+e^{y^a_j}+e^{y^b_j}+
e^{\gamma_{ab}+y^a_j+y^b_j}],
\label{recomsol}
\eeq
in which $\gamma_{ab}$ is the interaction function
(cf. \rref{intfunc}\ts)
$$
e^{\gamma_{ab}}=-{m_a^2+m_b^2-m_{a-b}^2-2m_a\sigma\over
{m_a^2+m_b^2-m_{a+b}^2+2m_a\sigma}},
$$
a complex function of $\sigma$.

First let us consider the initial form of the solution at $t=0$.
Suppose $m_a>\sigma_R>0$, then as in the previous case, we can make
the influence of the c-soliton as small as we wish
\beq
|\exp(\sigma x+x_0^{(b)}+{2\pi ib\over{n+1}}j)|<\epsilon \quad {\rm for}\quad
|x|<L,
\label{smallampb}
\eeq
by choosing $x^{(b)}_{0R}$ sufficiently large and negative.
Let us also require that
\beq
L\gg {1\over{m_a}}.
\eeq
Then at $x>L$ we have
$$
|e^{y^a_j}|\gg1
$$
and the argument of the logarithm in \rref{recomsol}\ts can be well
approximated by
\beq
e^{y^a_j}(1+e^{\gamma_{ab}+y^b_j}).
\label{recomampappr}
\eeq
So at $t=0$ and $x>L$
\beq
-i\beta\phi(x,0)\approx \sum_{j=0}^n\alpha_j\log[e^{y^a_j}
(1+e^{\gamma_{ab}+\sigma x+x_0^{(b)}+{2\pi ib\over{n+1}}j})].
\label{recompargu}
\eeq
Thus by the same argument as in the previous section,
$\phi(x,0)$ is regular for $x>L$.
And for $x<L$, the effect of the c-soliton is negligible
and $\phi(x,0)$, $x<L$ is given by the r-soliton solution,
which is regular.
Therefore $\phi(x,0)$ is regular everywhere. By continuity in $t$,
$\phi(x,t)$ is regular for sufficiently small $t$.

As $t$ increases, it becomes possible that at $x>L$
$$
1+e^{\gamma_{ab}+\sigma x-\lambda t+x_0^{(b)}+{2\pi ib\over{n+1}}j}
$$
vanishes. The solution is obtained by solving
\beq
\pmatrix{\sigma_R &-\lambda_R\cr
\sigma_I &-\lambda_I\cr}\pmatrix{x \cr t\cr}=\pmatrix{-x^{(b)}_{0R}
-(\gamma_{ab})_R\cr
(2m+1)\pi-{2\pi a\over{n+1}}j-x^{(b)}_{0I}-(\gamma_{ab})_I\cr}.
\label{mateqtwo}
\eeq
The singularity occurs at $t_M$ and the system is unstable.

As before
this instability can be naively ``understood" if we approximate
\rref{recomsol}\ts
in the region $x<L$,
\beq
-i\beta\phi(x,t)=\sum_{j=0}^n\alpha_j\ts e^{\sigma x-\lambda t +x_0^{(b)}+
{2\pi ia\over{n+1}}j+\gamma_{ab}}
\label{recompsolexp}
\eeq
which has $e^{-\lambda t}$, an exponentially growing or decaying
factor for complex $\lambda$.
It should be noted, however, that this approximation is not valid for $x>L$.

\medskip
As in the previous section we show the existence of solutions having lower
energies than the single r-solitons.
In other words these r-solitons do not have a ``mass gap".
Let
$\phi_r(x,t)$ be an explicit 1 r-soliton solution.
Consider the solution of the following initial value problem:
\beq
\phi(x,0)=\phi_r(x,0),\quad \partial_t\phi(x,0)=\partial_t\phi_r(x,0)+if(x),
\label{gapless}
\eeq
in which $f(x)$ is a real function and finite everywhere.
It is chosen to be orthogonal to
the real and imaginary parts of
$\partial_t\phi_r(x,0)$;
$$
\int_{-\infty}^\infty \partial_t\phi_{rR}^t(x,0)\cdot f(x)dx=0,\quad
\int_{-\infty}^\infty \partial_t\phi_{rI}^t(x,0)\cdot f(x)dx=0,
$$
and square integrable
$$
\int_{-\infty}^\infty f(x)^t\cdot f(x)dx=F>0.$$
The solution has a real total energy
$$
E=E_r-{1\over2}F,$$
in which $E_r$ is the total energy of the 1 r-soliton solution.

\section{Summary}
\setcounter{equation}{0}

As expected from the non-hermiticity of the Lagrangian,
the affine Toda field theory with a pure imaginary coupling constant
is found to be classically unstable:
It has many (almost all) solutions which develop  singularity after a
finite lapse of time;
Its energy is not positive definite;
Its potential is not bounded from below;
Small perturbation around soliton solutions does not necessarily
remain small as time passes.

 From these it seems that it is a long way to go to construct
a quantum theory of the non-hermitian affine Toda field theory with
a pure imaginary coupling constant,
although the theory has many beautiful classical solutions with remarkable
properties.
Thus the calculations of the quantum mass corrections to the solitons
are ill-founded.

\medskip

Although we have described many negative aspects of the affine Toda field
theory with imaginary coupling,  we are still fascinated by and interested in
the theory, especially  in the algebraic structure.
We believe that the solitons have exact and factorisable S-matrices
obeying non-trivial Yang-Baxter equations \cite{DGZ,Hod,GM}, reflecting the
integrability
of the theory.
However, due to the constraints from unitarity, we expect that these
S-matrices can make sense only for certain discrete values of the coupling
constant $\beta^2$,  at which, for example,  the corresponding conformal field
theories are known to be unitary \cite{HMBL}.
Therefore, the usual method of calculation in quantum field theory, the
perturbation calculation, is not  applicable to the solitons and their bound
states and no perturbative mass corrections to them.

\section*{Acknowledgments}
We thank E.\ts Corrigan, G.\ts Delius, T.\ts Hollowood, N.\ts MacKay and G.\ts
Watts for interesting  discussion.
R.\ts S thanks Department of Physics,
Kyung Hee University for hospitality, where part of this
work was done.
He is grateful to Japan Society for Promotion of Sciences
for financial support
which enabled him to visit Seoul.

\appendix

\section{Counter Example to ``hermiticity argument"}
\setcounter{equation}{0}
\renewcommand{\theequation}{A.\arabic{equation}}

In this appendix we give a simple counter example to the
``hermiticity argument" produced by Hollowood and Evans.
Let $V_2$ be a two dimensional complex vector space and
$f, g\in V_2$,
$$
f=\pmatrix{f_1\cr f_2\cr},\quad g=\pmatrix{g_1\cr g_2\cr},
$$
then the ordinary inner product is given by
$$
(f,g)=f^\dagger\cdot g=f_1\ccon g_1+f_2\ccon g_2.
$$
Next we choose the $Z_2$ symmetry matrix $M$ as
$$
M=\pmatrix{1&0\cr 0&-1\cr},\quad M^2=1,\quad M^t=M.
$$

Then a $2\times2$ matrix ${\cal D}$
\beq
{\cal D}=\pmatrix{a&b\cr -b\ccon &d\cr},\quad a, d :~{\rm real},
\quad b :~{\rm complex},
\label{Dexpform}
\eeq
satisfies the conjugation relation
\beq
{\cal D}^\dagger=M{\cal D}M.
\label{Drel}
\eeq
In fact, one can show easily that the above ${\cal D}$ is the most general
$2\times2$ matrix satisfying the relation \rref{Drel}.

According to Hollowood and Evans \cite{Hoc,Eva},
${\cal D}$ is ``hermitian" with
respect to a new ``inner product"
$$
\lang f,g\rang=f^\dagger\cdot Mg=f_1\ccon g_1-f_2\ccon g_2.
$$
Obviously $\lang f,f\rang$ is real but not positive (semi) definite.
The characteristic polynomial of ${\cal D}$ is
\Bear
{\rm det}\pmatrix{a-\lambda &b\cr -b\ccon &d-\lambda\cr}&=&
\lambda^2-(a+d)\lambda +ad
+|b|^2=(\lambda-\lambda_1)(\lambda-\lambda_2),
\label{charDeq}\\
\lambda_1+\lambda_2&=&a+d,\quad \lambda_1\lambda_2=ad+|b|^2.\NN
\Enar
Thus the eigenvalues are either both real or a complex conjugate pair
according to the sign of the discriminant
$$
(a+d)^2-4(ad+|b|^2)=(a-d)^2-4|b|^2.
$$
The eigenvalue problem reads
\Bears
ax+by&=&\lambda_1x,\\
-b\ccon x+dy&=&\lambda_1y.
\Enars
Assuming $b\neq0$, we get $y=(\lambda_1-a)x/b$ and
$$
\lang f,f\rang=|x|^2(1-|{\lambda_1-a\over b}|^2)=
{|x|^2\over{|b|^2}}(|b|^2-|\lambda_1-a|^2).
$$
If $\lambda_1$ is complex then $|\lambda_1-a|^2=(\lambda_1-a)(\lambda_2-a)$
and it is easy to see
$$
|b|^2-|\lambda_1-a|^2=0.
$$
Thus we find that the eigenvector belonging to a complex eigenvalue
 has a ``zero norm" (see \rref{nonrealrel})
\beq
\lang f,f\rang=0.
\label{zeronorm}
\eeq


\section{Symmetric and non-hermitian matrices}
\setcounter{equation}{0}
\renewcommand{\theequation}{B.\arabic{equation}}

In this appendix we show that the linear analysis   around a one
soliton solution proposed by Delius and Grisaru \cite{DGb}\ts
is incomplete with the
help of a simple example.

We consider a simple symmetric and non-hermitian $2\times 2$ matrix,
$$ {\cal D}= {\pmatrix {a+ib  & c+id \cr c+id  & g-ib }}, ~~a,b,c,d,g
:~{\rm real},~~ {\rm with}~~b(g-a)=2cd,
$$
\noindent so that, the coefficients of the characteristic polynomial
are real. The characteristic polynomial reads,
\beq \lambda^2-(a+g)\lambda +ag+b^2-c^2+d^2=0.  \label{charac}\eeq
The roots are $\lambda_1,\lambda_2 =A\pm B$, where,
$$ A=(a+g)/2 ,\quad {\rm and}$$
\beq B={\sqrt{(a-g)^2-4(b^2-c^2+d^2)}}/2. \label{roots}
\eeq

We show that if the eigenvalues are degenerate then the corresponding
eigenspace is one dimensional.
Moreover, in this case the eigenvector has ``zero norm'' in their
 symmetric ``inner product'' \rref{DGinerpro}.
So let us concentrate on the case of degenerate eigenvalues. In this
case, $B=0.$ The eigenvalue equation is given by

\beq (a+ib)x +(c+id)y={(g+a)\over 2}x,
\label{eigf}\eeq
\noindent so that
$$y={1\over {(c+id)}}[(g-a)/2 -ib]x,~~~c+id\neq 0.$$
\noindent Where we have taken $ f ={\pmatrix {x\cr y}},$  as the
eigenvector. Now let us examine the two cases viz $b=0$ and $b\neq 0$
separately:

Case i) $b=0$. In this case either  $c$ or $d$ is zero to maintain
the condition $b(g-a)=2cd$. So we have
$$ g-a = \pm 2 d\quad (\pm2ic)~~~~~~~~{\rm for}~~c=0\quad (d=0)$$
\noindent and consequently $y=\pm ix$.

Case ii) $b\neq 0$. In this case it is easy to see that vanishing
discriminant implies $b=\pm c,$ and in turn the condition $b(g-a)=2cd$
gives $(g-a)=\pm 2d$. So for this case again $y=\pm ix.$

For both  cases the eigenspace is spanned by only one vector.
Moreover the ``inner product'' defined by Delius and Grisaru
has vanishing norm,
\beq \llang f,f\rrang=f^t\cdot f=x^2+y^2= x^2-x^2 =0.
\label{inpro}
\eeq
Thus the ``complete set of orthogonal eigenfunctions''
of ${\cal D}$ does not exist.


\end{document}